\documentclass[final,3p,times,twocolumn]{elsarticle}

\usepackage{epsfig}

\usepackage{amssymb}
\usepackage{amsthm}

\usepackage{lineno}

\usepackage{natbib}
\usepackage{tabularx}
\usepackage{bm}
\usepackage{euscript}
\usepackage{graphicx}
\usepackage{color}
\usepackage{amsfonts}
\usepackage{exscale}
\usepackage{amsbsy}
\usepackage{textcomp}
\usepackage{dcolumn}
\usepackage{amsmath}
 \usepackage{xfrac}

\journal{Physica B}

\def\be{\begin{equation}}
\def\ee{\end{equation}}
\def\ba{\begin{eqnarray}}
\def\ea{\end{eqnarray}}

\def\C60{A$_x$C$_{60}$}

\begin{document}

\begin{frontmatter}



\title{Notes on Constraints for the Observation of Polar Kerr Effect in Complex Materials}


\author[su1,su2,su3]{Aharon Kapitulnik}


\address[su1]{Deapartment of Applied Physics, Stanford University, Stanford CA 94305, USA}
\address[su2]{Deapartment of Physics, Stanford University, Stanford CA 94305, USA}
\address[su3]{Stanford Institute for Materials and Energy Sciences, SLAC National Accelerator Laboratory, 2575 Sand Hill Road, Menlo Park, CA 94025, USA}

\begin{abstract}
While Kerr effect has been used extensively for the study of magnetic materials, it is only recently that its has shown to be a powerful tool for the study of more complex quantum matter.  Since such materials tend to exhibit a wealth of new phases and broken symmetries, it is important to understand the general constraints on the possibility of observing a finite Kerr effect.  In this paper we reviewed the consequences of reciprocity on the scattering of electromagnetic waves. In particular we concentrate on the possible detection of Kerr effect from chiral media with and without time-reversal symmetry breaking. We show that a finite Kerr effect is possible only if reciprocity is broken. Introducing the utilization of the Sagnac interferometer as a detector for breakdown of reciprocity via the detection of a finite Kerr effect, we argue that in the linear regime, a finite detection is possible only if reciprocity is broken. We then discuss possible Kerr effect detection for materials with natural optical activity, magnetism, and chiral superconductivity.
\end{abstract}


\end{frontmatter}


\section{Introduction}

The discovery by Bednorz and M$\rm \ddot{u}$ller (1986) of high temperature (high-Tc) superconductivity in copper oxides (cuprates) had an enormous impact on almost all aspects of research in quantum materials in general, and superconductivity in particular. Soon after this discovery, a whole range of novel superconducting, magnetic and metallic states were discovered in oxides and related systems. New pairing mechanisms associated with novel broken symmetries have been the highlight of the field of superconductivity, and the concept of unconventional superconductivity has emerged. Subsequent discoveries of other novel materials such as carbon nanotubes, graphene, and topological insulators, solidified the importance of quantum matter as a new paradigm in materials physics. Recent progress in the field both, theoretically and experimentally, suggests that local phenomena at the nanometer scale are the key to the novel behavior. The electrons have a very strong propensity to microscopically phase separate and to self-organize in patterns of lower-dimension. These observations also led to the emergence of new experimental techniques that are uniquely capable of probing these new aspects of matter. 

In this paper we concentrate on the Kerr effect as a probe for quantum states that exhibit violation of reciprocity. Our approach is to highlight the concept of reciprocity and discussing its role in nonlocal electrodynamics  even in the presence of dissipation. Pertaining to the possible observation of Kerr effect, we constraint it to violation of time reversal symmetry because of either spontaneous symmetry breaking such as magnetism or chiral superconductivity, or the application of an external magnetic field. As a consequence, our derivations demonstrate that a Kerr response from a reciprocal material-system such as gyrotropy due to natural optical activity, vanishes exactly. 

The organization of this paper is as follows. First we define the Polar Kerr effect (PKE), which is the primary subject of this paper. We then continue to discuss the concepts of reciprocity and time reversal symmetry (TRS) which are at the heart of understanding the possible observation of a finite PKE. We then apply these concepts to several important examples which include natural optical activity, magnetism and unconventional superconductivity.

\section{Polar Kerr effect}

It is customary to define the Kerr effect through the analysis of the state of polarization of light reflected from a magnetic solid, hence its MOKE (Magneto-optical Kerr effect) acronym. While the effect was discovered in 1877 by Rev. John Kerr \cite{Kerr1877}, its complete explanation had to wait for the introduction of quantum mechanics. Specifically, spin-orbit coupling and exchange splitting were shown to be necessary for a full understanding of the effect \cite{Argyres1955}. Of particular interest to us in this paper is the Polar Kerr Effect (PKE) which measures the rotation of a linearly polarized light reflected from a magnetized material at normal incidence.  

Assume a material that exhibits a ferromagnetic component of magnetization perpendicular to the surface of a sample. A linearly polarized light that is reflected from that surface will rotate exhibit a rotation of the polarization by a Kerr angle $\theta_K$ that reflects the fact that the indices of refraction for right ($+$) and left ($-$) circularly polarized light, which make up the linear polarization, are different. Using the convention in which the sense of circular polarization is determined with respect to a given axis (usually the z- axis), the Kerr angle will be determined by comparing the phase shifts of the reflected light of the two circular polarizations, $R_{++}$ and $R_{--}$.  Specifically
\be
\theta_K = \frac{1}{2}\left \{ \textrm{arg}[R_{++}]-\textrm{arg}[R_{--}]\right\}
\ee

The expression for Kerr effect in terms of the transition amplitudes can then be related to the dielectric function and the respective indices of refraction for the material \cite{Argyres1955}.
\be
\theta_K=\Im \left[ \frac{\tilde{n}_+ - \tilde{n}_-}{\tilde{n}_+\tilde{n}_- -1}\right].
\ee
Here $\tilde{n}_\pm$ are the complex indices of refraction for right and left circularly polarized light. We can define now the average index of refraction of the material as $\tilde{n} = (\tilde{n}_+ + \tilde{n}_-)/2$, and since in general, $|\tilde{n}_+-\tilde{n}_-|\ll |\tilde{n}| $, It can be shown that the above expression can be approximated by

\be
\theta_K= - \frac{4\pi}{\omega}\Im \left[ \frac{\sigma_{xy}}{\tilde{n}( \tilde{n}^2 -1)}\right].
\label{kerrsimple}
\ee
Obtaining the behavior of the off-diagonal terms of the electrical conductivity therefore can be used to calculate the Kerr effect in different materials \cite{Argyres1955}.

\section{Reciprocity}

In the following discussion, as well as appendices \ref{green} and \ref{scattering} we adopt the presentation of Tiggelen and Maynard \cite{Tiggelen1998}.  Assuming linear wave propagation according to the equation
\be
i\hbar \frac{\partial}{\partial t}\psi(\textbf{r},t)=\hat{\mathcal{H}}\psi(\textbf{r},t)
\label{scheq}
\ee
Here $\hat{\mathcal{H}}=\hat{\mathcal{H}}_0+\hat{V}$ where $\hat{\mathcal{H}}_0$ is the operator that describes free space wave propagation and $\hat{V}$ describes the scatterer's potential.

Reciprocity is defined by applying an anti unitary operator to the physical process. An anti unitary operator $\hat{K}$ applied to a linear combination of two states, $\alpha$ and $\beta$ yields
\be
\hat{K}(a\psi_\alpha + b\psi_\beta)=a^*\hat{K}\psi_\alpha + b^*\hat{K}\psi_\beta
\ee
which is called anti linearity property. Here ``$*$" denotes complex conjugation.  Also, the inverse of $\hat{K}$ coincides with its adjoint $\hat{K}^{-1}=\hat{K}^\dag$, and the scalar product satisfies
\be
\left(\hat{K}\psi_\alpha,\hat{K}\psi_\beta \right)=\left(\psi_\alpha,\psi_\beta \right)
\ee
Finally, it can be shown that any antiunitary $\hat{K}$ can be written in the form
\be
\hat{K}=\hat{U}\hat{C}
\ee
where $\hat{U}$ is a unitary operator and $\hat{C}$ is a conjugation operator. Applying $\hat{C}$ to the Schr\"oedinger equation yields the time reversed solution to that equation.

Let us assume that the free hamiltonian commutes with the anti unitary operator $\hat{K}$ such tat 
\be
\hat{K}\hat{\mathcal{H}}_0\hat{K}^{-1}=\hat{\mathcal{H}}_0
\ee
But, due to absorption it does not commute with the full Hamiltonian, but rather
\be
\hat{K}\hat{V}\hat{K}^{-1}=\hat{V}^\dag
\label{rec1}
\ee
such that 
\be
\hat{K}\hat{\mathcal{H}}\hat{K}^{-1}=\hat{\mathcal{H}}^\dag
\label{rec2}
\ee

Equations~\ref{rec1} and~\ref{rec2} are called the reciprocity condition, and $\hat{K}$ is the reciprocity operator for the system. The above definition becomes clear when $\hat{K}$ is applied to the Green's functions, yielding the familiar reciprocity theorem for the GreenÕs operator.
\be
\hat{K}G_E^{\pm}\hat{K}^{-1}=G_E^{\pm \dag}
\label{recg}
\ee

Let us first consider the case of elastic scattering such that $\phi_{E_{\alpha}}\rightarrow \phi_{E_{\beta}}$, but $E_\alpha = E_\beta = E$. Applying $\hat{K}$ to the scattered wavefunction yields
\be
\hat{K}\psi_{E_{\alpha}}^{\pm}=\psi_{E_{\alpha}}^{T\mp}
\ee
Which indeed confirms the fact that $\hat{K}$ reverses the scattering process to the reversed one. Thus, the transition amplitudes calculated above will satisfy
\be
\langle \beta | \hat{T}  |\alpha \rangle = \left( \phi_{E_\beta}, \hat{V}\psi_{E_\alpha}^+ \right) = \left([ \hat{K}\psi_{E_\alpha}^{T-}], \hat{V}[\hat{K}\phi_{E_\beta}] \right)\equiv \langle \bar{\beta} | \hat{T}  |\bar{\alpha} \rangle
\ee
This is the reciprocity theorem for the transition amplitude as it relates the scattering process to its reversed one.

\section{Time Reversal Symmetry}

In the Schr\"oedinger formulation of the wave equation without spin-consideration (appropriate for our discussion of radiation), the time-reversal operation $\hat{\mathcal{T}}$ may be defined as the complex conjugation operator $\hat{C}$, in the position representation \cite{Sakurai1985}.  The condition for time reversal symmetry without absorption occurs when
\be
\hat{\mathcal{T}}\hat{\mathcal{H}}\hat{\mathcal{T}}^{-1}=\hat{\mathcal{H}}
\ee
An important example that will be used later is that of a circularly polarized plane wave with wavevector $\textbf{k}$ and with circular polarization $\sigma = \pm$ for right and left circularly polarized light respectively. Applying $\hat{C}$ we have
\be
\hat{C}|\textbf{k}, \sigma \rangle = |-\textbf{k},\sigma \rangle
\ee
Note that $\sigma$ does not change sign as helicity is invariant under the time reversal operation.

If absorption is present, applying the time reversal operator yields
\be
\hat{\mathcal{T}}\hat{\mathcal{H}}\hat{\mathcal{T}}^{-1}=\hat{\mathcal{H}}^*
\ee
Note that even if time reversal is not a symmetry - e.g., in the presence of if absorption, the reciprocity theorem can be offered by the time reversal operator, as long as it fulfills the reciprocity conditions as in equations~\ref{rec1} and~\ref{rec2}. While this statement emphasizes that time reversal symmetry is not the same as reciprocity, it can be a reciprocity operator even if it is not the symmetry of the system.

We note that an external magnetic field disrupts the above relation as it requires to change the direction of the magnetic field for reciprocity to hold
\be
\hat{\mathcal{T}}\hat{\mathcal{H}}(\textbf{H})\hat{\mathcal{T}}^{-1}=\hat{\mathcal{H}}(-\textbf{H})
\label{field}
\ee
Thus we conclude that  time reversal symmetry is broken in an external magnetic field.

Finally we note that as we previewed in the introduction, time reversal symmetry is sometimes confused with reciprocity. Obviously, in the presence of absorption time reversal symmetry is not respected, while, as we have argued above, reciprocity is still intact.\\

\section{Constraints on the possible observation of Kerr effect}
\subsection{Motivation and general remarks }
The recent discovery that the occurrence of charge order in several compounds of high-Tc superconductors coincide with the previously detected onset of Kerr effect in the same compounds (see e.g. Wu {\it et al.} \cite{Wu2014})  prompted a renewed discussion on the possible detection of Kerr rotation from optically active gyrotropic materials. In particular,  Hosur {\it et al.} \cite{Hosur2013} and then Mineev \cite{Mineev2010}, proposed that the Kerr effect arises from chiral charge density wave ordering.  In its core, this discussion revolves around whether Kerr rotation is allowed within linear response in materials which break inversion ( $\mathcal{I}$) and all mirror symmetries, but preserve time-reversal symmetry. 

Pertaining to the high-Tc cuprates, such proposals already appeared in the early days of this field to describe results obtained in search for signatures of "anyon superconductivity" \cite{Halperin1992} in these materials \cite{Arfi1992}.  The invention of a Modified Sagnac Interferometer to test for the observed effects \cite{Spielman1990} and subsequently refute the anyons proposal, \cite{Spielman1990,Spielman1992} was a significant advancement in this debate since it was the first time that an apparatus was constructed to yield positive Kerr effect results \emph{only} in the presence of broken time reversal symmetry (TRSB) \cite{Halperin1992}.    Dodge {\it et al.} \cite{Dodge1996} further studied the symmetry properties of the  Sagnac interferometer as an apparatus that respects reciprocity, while allowing for a detectable signal for TRSB effects.  Moreover, Dodge {\it et al.} \cite{Dodge1996} also concluded that materials with natural optical activity without TRSB will not yield a finite Kerr effect.  Therefore, the association of the recent Kerr results on high-Tc cuprates with optical activity without time reversal symmetry breaking (TRSB) should have been deemed wrong already at their inception \cite{MineevErratum2014,Fried2014b}. 

A primary reason for the continuing controversy on the possible Kerr effect from non-TRSB gyrotropy has been a confusion in distinguishing between the concepts of time reversal symmetry and reciprocity, and consequently in an error when applying linear response theory to the Kerr effect from gyrotropic materials.  Such an error typically appears when solving the macroscopic Maxwell equations with a non-TRSB gyrotropic term by calculating the relevant Fresnel coefficients, yielding a non-zero Kerr effect.  As is often done in electrodynamics, reciprocity is revealed as time-reversal symmetry breaking  by demonstrating invariance under the interchange of source and detector \cite{LL1960}.  However, systems that exhibit absorption clearly violate time reversal symmetry, but may respect reciprocity. Here we define reciprocity in linear response systems as the self-transpose properties of the scattering matrix and its related response functions given in terms of the dielectric function and magnetic permeability.  In the case of light scattering, which is the subject of the present paper,  can be shown to hold even in the presence of absorption depending on the chosen basis of polarization.

\subsection{Reciprocity Considerations}

The Kerr effect defined above is intimately connected to the concept of reciprocity. Since Kerr effect is defined through an asymmetry of reflection amplitudes of light from a given material, it is most naturally considered within the general theory of scattering.  Indeed in Appendix~\ref{green} and \ref{scattering} we outline the theory of scattering of light from a general dielectric where the symmetry of the scattering process, including the symmetry of the dielectric material itself are contained in the scattering potential $V_{ij}(\textbf{r},\textbf{r}^\prime;\omega)$

 Reciprocity as defined in equation~\ref{reciprocity} requires that the scattering potential matrix $V_{ij}$ satisfies
\be
V_{ij}(\textbf{r},\textbf{r}^\prime;\omega) = V_{ji}(\textbf{r}^\prime,\textbf{r};\omega)
\ee
which, in turn requires that (see also Xie {\it et al.} \cite{Xie2009})
\be
\varepsilon_{ij}(\textbf{r},\textbf{r}^\prime;\omega) = \varepsilon_{ji}(\textbf{r}^\prime,\textbf{r};\omega)
\label{dielectric}
\ee
which is nothing but Onsager reciprocity relation \cite{Onsager1931} for the dielectric tensor. Note that we did not make any assumption here about absorption, which may still be embedded in $\varepsilon$. This suggests that we can use the general formalism of Onsager's relations to predict the outcome of a Kerr effect experiment. 

At this point it is important to note that while it is convenient to use the concept of reciprocity to understand Kerr effect,  the use of fluctuation-dissipation theorem, which is another principal result of linear response theory (that is, the study of irreversible thermodynamics associated with linear processes that are referred to the equilibrium state of a system)  yields an identical result for the symmetry of the dielectric function in Eqn.~\ref{dielectric}.  Intuitively, both results, Onsager reciprocity relations and fluctuation-dissipation theory rely on the assumption that near equilibrium, macroscopic response and decay process occur in the same manner as the decay of equilibrium fluctuations. 

To measure the Kerr effect we send circularly polarized light onto a material and examine the amplitude and polarization of the reflected light.  At normal incidence, which is the configuration of a Polar Kerr Effect (PKE) We compare the transition amplitude of a wave at state $|\textbf{k},+\rangle$ which is backscattered into a state $|-\textbf{k},+\rangle$, with the transition amplitude of a wave at state  $|\textbf{k},-\rangle$, which is back-scattered into a state $|-\textbf{k},-\rangle$.  Here again, we are using the convention that the sense of circular polarization is determined with respect to a given axis (usually the $z$-axis). In this case a reflected light will have the same sense of polarization which means that the handedness relative to the direction of propagation changes.These two amplitudes are basically the two reflection coefficients\ba
\langle \textbf{k}, + |\hat{T}(\omega) | -\textbf{k}, + \rangle \nonumber \equiv R_{++}\\
\langle \textbf{k}, - |\hat{T}(\omega) | -\textbf{k}, - \rangle \equiv R_{--}
\ea
Since the sense of circular polarization is fixed with respect to a given axis, but is opposite when attached to the direction of propagation, it is the symmetry of the transition amplitude matrix that dictates whether reciprocity is maintained. 
  If reciprocity holds,
\be
\langle \textbf{k}, + |\hat{T}(\omega) | -\textbf{k}, + \rangle \nonumber =\langle \textbf{k}, - |\hat{T}(\omega) | -\textbf{k}, - \rangle 
\ee
implying that the Kerr effect is zero since the two reflection amplitudes are equal $R_{++}=R_{--}$, and
\be
\theta_K = \frac{1}{2}\left \{ \textrm{arg}[R_{++}]-\textrm{arg}[R_{--}]\right\}=0
\ee

\subsection{Natural Optical Activity}

Of particular interest to us is the situation of a material which exhibits natural optical activity \cite{HosurErratum2014}. To test whether a finite Kerr effect is possible for such a material we need to look at the structure of the dielectric tensor $\varepsilon_{ij}(\textbf{r},\textbf{r}^\prime;\omega)$, requiring that it will exhibit reciprocity. The electric displacement vector is now given in terms of the non-local dielectric tensor and the electric field:
\be
D_i(\textbf{r};\omega)=\int d^3r^\prime \varepsilon_{ij}(\textbf{r},\textbf{r}^\prime;\omega) E_j(\textbf{r}^\prime;\omega)
\ee

The symmetry of the dielectric tensor as a result of reciprocity which is expressed in equation~\ref{dielectric} can be expressed as follows. Suppose for an applied electric field $\textbf{E}_1(\textbf{r})$ the response electric displacement is $\textbf{D}_1(\textbf{r})$, while for an applied electric field $\textbf{E}_2(\textbf{r})$ the response electric displacement is $\textbf{D}_2(\textbf{r})$. Reciprocity then requires that  \cite{LL1960}
\be
\int d^3r  \textbf{D}_1\cdot\textbf{E}_2 = \int d^3r  \textbf{D}_2\cdot\textbf{E}_1
\label{llreciprocity}
\ee
This reciprocity requirement should hold for any pair of electric field and electric displacement is given through the solution of the full Maxwell equations including all boundaries. However, while the bulk response of a material with natural optical activity is well established as:
\be
D_i(\textbf{r}) = \varepsilon_{ij}^0E_j(\textbf{r})+\gamma_{ijk}\frac{\partial E_j(\textbf{r})}{\partial x_k},
\label{gyrobulk}
\ee
the boundary conditions at the interface between that material and free space have been a subject of great ambiguity \cite{Vinogradov2002}. To resolve this ambiguity we apply the above bulk constitutive relation (Eqn.~\ref{gyrobulk}) to the reciprocity condition (Eqn.~\ref{llreciprocity}). The result is the following requirement for any pair $[\textbf{E}_1(\textbf{r}), \textbf{D}_1(\textbf{r})]$ and $[\textbf{E}_2(\textbf{r}), \textbf{D}_2(\textbf{r})]$:
\ba
\int d^3r \gamma_{ijk}(\textbf{r};\omega) \frac{\partial E_{1j}(\textbf{r})}{\partial x_k}E_{2i}(\textbf{r}) = \nonumber \\ 
\int d^3r \gamma_{ijk}(\textbf{r};\omega) \frac{\partial E_{2j}(\textbf{r})}{\partial x_k}E_{1i}(\textbf{r}) 
\label{recip7}
\ea
Integrating by parts the left hand-side, and using the fact that $\gamma_{ijk}=-\gamma_{jik}$,  the principal terms indeed cancel, and we are left with two excess terms that do not necessarily vanish (where the sign of the terms is important)
\be
\int d^3r \frac{\partial}{\partial x_k}\left[ E_{2i}\gamma_{ijk}E_{1j}\right] - \int d^3r E_{1j}E_{2i}  \frac{\partial \gamma_{ijk}}{\partial x_k}
\label{recip6}
\ee

To understand these excess terms, we first notice that the integral on the left is an integral of $\nabla \cdot [\textbf{E}_1\times (\overset{\text{\tiny$\leftrightarrow$}}{\gamma}\cdot\textbf{E}_2)]$ and thus can be transformed into a surface integral. Integrating over all space, the surface integral is zero at infinity.  To avoid singularities, we can assume that $\gamma$ changes smoothly over some distance (much smaller than the wavelength of light) from the surface.  Now the surface integral is zero everywhere including the interface, and we are left with the second term that includes the gradient of $\gamma$ which is finite within a thin surface layer. This term appears with a negative sign, and it is therefore easy to see that if we split it into two, each half could cancel an equivalent term with opposite sign that we now deem missing from the constitutive relation of Eqn.~\ref{gyrobulk}, and thus from the reciprocity condition in Eqn.~\ref{recip7}.  Since $\textbf{E}_1$ and $\textbf{E}_2$ are arbitrary,  symmetry implies that this is the only possible choice to split the excess integral.  It is easy to show that the above result also holds if $\gamma$ changes  abruptly at the interface (as a $\theta$-function). In this case we define $\gamma$ at the interface to be the average value (i.e. one-half) of $\gamma$ between the vacuum (where $\gamma$ vanishes) and the material (where $\gamma$ is finite), and use the antisymmetry property of $\gamma_{ijk}$ \cite{Vinogradov2002}. 

We therefore conclude that the correct constitutive relation for a material with natural optical activity is:
\be
D_i(\textbf{r}) = \varepsilon_{ij}^0E_j(\textbf{r})+\gamma_{ijk}\frac{\partial E_j(\textbf{r})}{\partial x_k}+\frac{1}{2}\frac{\partial \gamma_{ijk}(\textbf{r})}{\partial x_k}E_j(\textbf{r})
\label{gyro}
\ee
If this relation is used to calculate the reflection amplitudes within the Fresnel equations formalism one can verify that indeed $\theta_K=0$ for a material with natural optical activity, irrespective of dissipation in the system, as we obtained from the general symmetry considerations.

\subsection{Magneto-optical effects}

Having proven that the Kerr effect vanishes for a gyrotropic system where gyrotropy originates from natural optical activity, we turn to the other possibility for gyrotropy which originates from time-reversal symmetry breaking either by an external magnetic field $\textbf{H}$, or a finite magnetization $\textbf{M}$. Starting with a static external magnetic field, $\textbf{H}$, we apply the time reverse operator, and in the presence of absorption, the scattering transition matrix satisfies
\be
\hat{\mathcal{T}}\overset{\text{\tiny$\leftrightarrow$}}{T}(\textbf{r},\textbf{r}^\prime;\omega,\textbf{H})\hat{\mathcal{T}}^{-1} = \overset{\text{\tiny$\leftrightarrow$}}{T}^*(\textbf{r}^\prime,\textbf{r};\omega,-\textbf{H})
\ee
yielding the following relation
\be
\langle \textbf{k},\sigma |\hat{T}(\omega,\textbf{H}) | \textbf{k}^\prime,\sigma^\prime \rangle = \langle -\textbf{k}^\prime,\sigma^\prime  |\hat{T}(\omega,-\textbf{H}) |- \textbf{k},\sigma \rangle
\label{reciprocityt}
\ee
which demonstrates that an external magnetic field breaks reciprocity in light propagation. This in turn gives
\be
\langle \textbf{k}, + |\hat{T}(\omega,\textbf{H}) | -\textbf{k}, + \rangle  \neq \langle \textbf{k}, - |\hat{T}(\omega,\textbf{H}) | -\textbf{k}, - \rangle 
\ee
implying that a finite Kerr effect is now possible since the two reflection amplitudes are not equal $R_{++}\neq R_{--}$, and
\be
\theta_K = \frac{1}{2}\left \{ \textrm{arg}[R_{++}]-\textrm{arg}[R_{--}]\right\} \neq 0
\ee
Similar expressions can be written for a spontaneous magnetization $\textbf{M}$, although here we need to be careful about its possible spatial dependence.

\subsection{Unconventional superconductors}

The discovery of superconductivity in heavy-electron systems (first in CeCu$_2$Si$_2$ \cite{Steglich1979}) gave birth to the field of ``unconventional superconductors" \cite{Norman2011},  that is, superconductors that do not conform to the conventional BCS theory \cite{BCS1957} or its immediate variants (such as Eliashberg theory  \cite{Eliashberg1960} or the Bogolubov-de-Gennes theory\cite{BgdG}.) While attraction and pairing are still believed to be key ingredients, unconventional superconductivity may have a different origin than electron-phonon mechanism, and together with strong correlations that may exist in such materials may result in a pairing state of higher angular momentum.  While $s$-wave superconductors inherently respect time reversal symmetry (see e.g. Anderson's theorem \cite{Anderson1959}), unconventional superconductors can be found in forms that break it (we call these ``chiral"). Indeed, soon after the discovery of the layered-perovskite superconductor Sr$_2$RuO$_4$ \cite{Maeno1994},  it was predicted to be an odd-parity superconductor \cite{Rice1995,Baskaran1996} with a gap function that satisfies $\Delta(\textbf{p}) \propto p_x \pm ip_y $. The two states with the $\pm$ sign break TRS in an ``Ising spin fashion," giving rise to an asymmetry in the response of the system to electromagnetic excitation. Similar to a state with finite magnetization discussed above, one expect an asymmetry in the optical response of right/left circularly polarized light.  Thus, magneto-optic-like effects could be the obvious tests for TRS-breaking in unconventional superconductors.

Using the example of a $p_x \pm ip_y $ superconductor, it is easy to see however that while indeed an asymmetry may exist for the response of right and left circularly polarized light, the resulting Kerr effect of a single band, pure system (that is, translational symmetry is respected) is identically zero \cite{Roy2008,Lutchyn2008}.  A finite Kerr effect can be observed if either the material breaks translational symmetry \cite{Goryo2008,Lutchyn2009}, or it has more than one band crossing the Fermi energy where superconductivity originates, and inter band coupling is present \cite{Taylor2012}. Calculating $\sigma_{xy}$ including the inter-band transitions which carry the information on TRSB,  Eqn.~\ref{kerrsimple}, can then be used to obtain a finite expression for the Kerr effect. 

Indeed, a finite Kerr effect was measured for a number of unconventional superconductors including Sr$_2$RuO$_4$ \cite{Xia2006b} and UPt$_3$ \cite{Schemm2014}. In both cases high quality samples with RRR exceeding several hundreds were used. Thus, it is reasonable to assume that the finite Kerr effect observed originates from the multi-band nature of both materials.

\section{Kerr Effect Measurements with the Sagnac Interferometer}

The Sagnac interferometer is a two-beam polarization ring interferometer that was first proposed to detect mechanical rotation. In that standard configuration, which can be realized using bulk-optics components or a fiber-optic loop, a linearly-polarized beam of light is split using a beam-splitter into two counter propagating beams, enclosing a finite area. The two beams are brought back through the same beam splitter to interfere at the detector. In the presence of a finite rotation, a phase shift is detected that is proportional to the area enclosed by the loop and the projection of the angular velocity on the normal to the area. Implementing the detector at the same side as the source ensures that the Sagnac interferometer is fully reciprocal in the absence of rotation. In fact, in the regime where the medium through which the light propagates is linear -- that is, its refractive index does not vary with optical field strength -- the Sagnac interferometer will measure a zero phase shift unless a non-reciprocal component is inserted in the loop, or, in some cases if time-varying physical effects are present (for a recent discussion, see ref.~\cite{Culshaw2006}).

The Stanford group has developed a series of Sagnac interferometer systems and has studied their performance \cite{Spielman1990,Spielman1992,Dodge1996,Kapitulnik1994,Xia2006,Fried2014} as sensitive instruments for the detection of  non-reciprocal effects in solid-state systems. Most importantly, these interferometers, which were all based on fiber-optic waveguides, were carefully characterized for their performance to reject reciprocal effects. In general the interferometers can be divided into two groups. The finite-loop interferometers, for which a comprehensive discussion is given by Dodge {\it et al.} \cite{Dodge1996}, including an analysis of the interferometer's performance to measure only non-reciprocal effects. In fact the reciprocity condition of eqn.~\ref{llreciprocity} is given in eqn. (3) in ref.~\cite{Dodge1996}, and shown to hold for that system.

The second class of Sagnac interferometers, first reported by Xia {\it et al.},  \cite{Xia2006} exhibits a Sagnac loop which does not enclose any physical area. In that version, which was recently used to study the Kerr effect in high-Tc superconductors, \cite{Xia2008} a single polarization maintaining fiber (PMF) is used to constitute a zero-area loop in which two counter propagating beams use the two polarization states of the fiber as a waveguide. Emerging out of the fiber-strands, the two linear polarizations go through a quarter waveplate, and the resulting two circular polarizations interact with the material from which they are reflected. The two reflected beams go back through the quarter waveplate into the fiber strand and interfere at the detector.  Placing the detector next to the source with a directional circulator determining the emerging and returning beams, and with a reciprocal mirror in place of a sample, this interferometer is fully reciprocal and the output is identically zero (except for instrumental offset that is fully characterized before measurements commence.)  Operationally, by  modulating the two counter-propagating beams using an electro-optic phase modulator, and locking-in to the modulation, the ratio of first and second harmonic outputs yields the desired phase shift $\varphi_{nr}$ which is finite in the presence of any non-reciprocal effect along the optical circuit. A comprehensive discussion of the performance of a zero-area loop Sagnac interferometer is given in ref.~\cite{Fried2014}. While a reciprocal mirror returns a zero phase shift, if a non-reciprocal sample is measured, a finite phase shift $\varphi_{nr}$ will be measured at the detector, which is given by the difference 
\be
 \varphi_{nr} = \textrm{arg}[R_{++}]-\textrm{arg}[R_{--}] =2\theta_K
\ee

Thus, the Sagnac interferometer can be viewed as a detector for non-reciprocal effects in the material that is being examined. If a finite Kerr effect is measured in linear response, it means that the sample breaks reciprocity, either because a finite magnetic field is applied, or  spontaneous time-reversal symmetry breaking occurs, or time dependence is present to alter the properties of the investigated material in a fashion that is not averaged out within our detection scheme. Of course, in the non-linear regime, no reciprocity considerations are necessary to obtain a finite phase shift.\\

 \section{Conclusions}
In this paper we reviewed the possible ways to obtain a finite Kerr effect in different types of materials. In particular we highlighted the consequences of reciprocity on the scattering of electromagnetic waves, and discussed the possible detection of Kerr effect from chiral media with and without time-reversal symmetry breaking. We showed that a finite Kerr effect is possible only if reciprocity is broken. Finally, we discussed the utilization of the Sagnac interferometer as a detector for breakdown of reciprocity via the detection of a finite Kerr effect. 

\appendix
\section{The Green's Function Approach to Scattering}
\label{green}
Let us come back to the distinction between time reversal symmetry and reciprocity. In a scattering process time reversal symmetry will refer to the time evolution of the process, while reciprocity will refer to the scattering process. Applying the time reversal operator to the Schr\"oedinger equation is equivalent to applying the anti-linear operator of complex conjugation. However, care needs to be taken with respect to Maxwell equations that describe macroscopic media with constitutive equations that may include absorption.  However, the two problems are similar in that they can be formulated using the general theory of Green's functions as applied to waves scattering. Let us review this formalism. Assuming linear wave propagation according to the equation
\be
i\hbar \frac{\partial}{\partial t}\psi(\textbf{r},t)=\hat{\mathcal{H}}\psi(\textbf{r},t)
\label{scheq}
\ee
Here $\hat{\mathcal{H}}=\hat{\mathcal{H}}_0+\hat{V}$ where $\hat{\mathcal{H}}_0$ is the operator that describes free space wave propagation and $\hat{V}$ describes the scatterer's potential. Ampere and Faraday's equation, together with the usual constitutive relations of Ohm's law for the conductivity, and linear dependence on the dielectric constant and magnetic permeability are already in the Schr\"oedinger's formalism with a first-order time derivative 
\ba
\mu\frac{\partial \textbf{H}}{\partial t}=-c\nabla \times \textbf{E} \nonumber \\
\varepsilon\frac{\partial \textbf{E}}{\partial t}=-c\nabla\times \textbf{H} -4\pi \sigma \textbf{E}
\ea
Thus, to complete the analogy we identify the particle's wavefunction in the Schr\"oedinger's picture as the 6-components vector of the electromagnetic field \cite{Tiggelen1998, Deak2012}. 
\ba
\psi(\textbf{r},t) = \begin{pmatrix}
     \textbf{E}   \\
  \textbf{B}
\end{pmatrix}
\ea

The Hamiltonian operator will now be a $6 \times 6$ matrix. For example, the free-wave propagation operator will be
\ba
\hat{\mathcal{H}}_0=\begin{pmatrix}
     0 & \hat{e}\cdot\textbf{p}   \\
   - \hat{e}\cdot\textbf{p}   &  0
\end{pmatrix}
\label{freeh}
\ea
Here $\hat{e}$ is the L\'evi-Civita tensor of rank three,  $\textbf{p}=\frac{\hbar}{i}\nabla$ is the momentum operator, and $\hat{e}\cdot\textbf{p}=e_{ijk}p_k$. In matter, that is, in the presence of a scatterer, we will add a potential $\hat{V}$ which is described by the dielectric constant, and induced currents $\textbf{j}$, hence electrical conductivity.  Maxwell equations are then written in the form of equation~\ref{scheq} \cite{Tiggelen1998, Deak2012}.

We can now use the full power of scattering theory using Green's functions approach, such as the Lippmann-Schwinger formalism, to look at the symmetry properties of the scattering problem, especially with respect reciprocity. The time evolution of the wavefunction from time $t=0$ will be given by
\be
\psi(\textbf{r},t)=e^{-i\hat{\mathcal{H}}t/\hbar}\psi(\textbf{r},0)
\ee
For the case of no absorption, $\sigma(\textbf{r})=0$ and $\varepsilon(\textbf{r})=\varepsilon^*(\textbf{r})$, which lead to $\hat{V}=\hat{V}^{\dag}$, and hence  $|\psi(\textbf{r},t)|^2$ is a conserved quantity which in the case of Maxwell equations lead to conservation of the electromagnetic energy in time. 

The Stationary Green's operators are given by
\be
G_E^{\pm}=\left[ E - \hat{\mathcal{H}} \pm  i\epsilon \right]^{-1}
\ee
with $\epsilon \rightarrow 0$ and the plus or minus signs define the  retarded and advanced Green's operator respectfully. The energy eigenstate in the electromagnetic case is given by the frequency of the monochromatic light $E=\hbar \omega$.

Following standard notations (see e.g. \cite{Schiff1968}), we can write the wavefunction that grows in time out of the free particle wavefunction (hence the plus sign) as
\ba
\psi_{E_\alpha}^+(\textbf{r},t)=\phi_{E_\alpha} + G_E^+\hat{V}\phi_{E_\alpha} \nonumber \\
\psi_{E_\alpha}^{T+}(\textbf{r},t)=\phi_{E_\alpha} + G_{E}^-\hat{V}^\dag\phi_{E_\alpha}
\ea
Here the superscript $T$ is used for the transposed wavefunction, and $\phi_{E_\alpha}$ is the free propagation field solving the wave of the free Hamiltonian $\hat{\mathcal{H}}_0$, with an eigenenergy $E_{\alpha}$. Note that since we did not assume $\hat{V}$ to be self-adjoint, absorption effects are allowed to be incorporated. The transition operator $\hat{T}$ operating between free propagation initial state $\alpha$ and final state $\beta$ yields a transition amplitude
\be
\langle \beta | \hat{T}  |\alpha \rangle = \left( \phi_{E_\beta}, \hat{V}\psi_{E_\alpha}^+ \right) = \left( \psi_{E_\beta}^{T-}, \hat{V}\phi_{E_\alpha} \right).
\label{transition}
\ee

\section{Scattering of the Electric Field}
\label{scattering}

Establishing the framework of the discussion related to reciprocity and time reversal symmetry, we turn to an experimental situation in which we send a monochromatic electric field at frequency $\omega$ at a sample, and detect the scattered electric field far away from the scatterer. To achieve an equation for the electric field only we need to combine the Maxwell equations obtained in Eqn.~\ref{scheq}. This can easily be done and the result is a $3\times3$ matrix equation for the electric field only. For example, the $6\times6$ matrix equation for free space waves that uses $\hat{\mathcal{H}}_0$ in equation~\ref{freeh} yields the following equation for the electric field
\be
\left\{ -\nabla_T^2+ [1-\overset{\text{\tiny$\leftrightarrow$}}{\varepsilon} (\textbf{r})] \frac{\omega^2}{c^2}\right \} \cdot \textbf{E}(\textbf{r})= \frac{\omega^2}{c^2} \cdot \textbf{E}(\textbf{r})
\label{electric}
\ee
where we used the common value $\mu=1$ for the magnetic permeability (we will use this value for the rest of this manuscript as it applies for all materials at optical frequencies \cite{LL1960}, and denoted by $\overset{\text{\tiny$\leftrightarrow$}}{\varepsilon} (\textbf{r})$ the dielectric tensor. The operator $\nabla_T^2$ is the transverse projection of the laplacian. The momentum operator is written explicitly as a gradient, while the energy of the wave satisfies  $E=\hbar \omega$. The scattering potential is easily identified with
\be
[1-\overset{\text{\tiny$\leftrightarrow$}}{\varepsilon} (\textbf{r})] \frac{\omega^2}{c^2} \equiv \hat{V}(\textbf{r};\omega)
\label{scatter}
\ee
We note that this result will remain the same in the presence of a finite conductivity since this can be absorbed in the definition of the dielectric tensor via 
\be
\overset{\text{\tiny$\leftrightarrow$}}{\varepsilon} (\textbf{r})=\varepsilon_\infty \overset{\text{\tiny$\leftrightarrow$}}{I} +\frac{4\pi i }{\omega} \overset{\text{\tiny$\leftrightarrow$}}{\sigma} (\textbf{r})
\ee
 This expression can then be used to define the transition operator introduced in equation~\ref{transition}. In fact, using the operator form of equation~\ref{electric} we can write the transition operator as a Born series \cite{Schiff1968}
\be
\hat{T}(\omega)=\hat{V}(\omega) + \hat{V}(\omega) G_\omega \hat{V}(\omega) + \cdots
\ee
where
\be
G_\omega=G_\omega(\textbf{r}) =\left[ \frac{ \omega^2}{c^2} +\nabla_T^2 +i\epsilon \right]^{-1}
\ee
is the retarded  Green's function for the transverse waves operator $\frac {\omega^2}{c^2}+ \nabla_T^2$.

If in addition we have a non-local dielectric function this equation will be modified to yield a different scattering potential, but the structure of the above formulation will remain the same.  The scattering potential in this case is a simple generalization of Eqn.~\ref{scatter} (remember that $\mu=1$)
\begin{multline}
V_{ij}(\textbf{r},\textbf{r}^\prime;\omega) \equiv \\
- \frac{\partial}{\partial x_i}\frac{\partial}{\partial x^\prime_j}\delta(\textbf{r}-\textbf{r}^\prime)-\frac{\omega^2}{c^2} [\varepsilon_{ij}(\textbf{r},\textbf{r}^\prime;\omega)-\delta_{ij}\delta(\textbf{r}-\textbf{r}^\prime)]
\label{recip5}
\end{multline}
Equation~\ref{electric} is now an integral equation reading
\be
\nabla^2\textbf{E}(\textbf{r};\omega)+\frac{\omega^2}{c^2}\textbf{E}(\textbf{r};\omega)=\int d^3r^\prime \overset{\text{\tiny$\leftrightarrow$}}{V}(\textbf{r},\textbf{r}^\prime;\omega)\cdot\textbf{E}(\textbf{r}^\prime;\omega)
\ee
while the Born series reads now
\begin{multline}
 \overset{\text{\tiny$\leftrightarrow$}}{T}(\textbf{r},\textbf{r}^\prime;\omega)= \overset{\text{\tiny$\leftrightarrow$}}{V}(\textbf{r},\textbf{r}^\prime;\omega)+\\
 \int d^3r_1d^3r_2 \overset{\text{\tiny$\leftrightarrow$}}{V}(\textbf{r},\textbf{r}_1;\omega)G(\textbf{r}_1-\textbf{r}_2;\omega) \overset{\text{\tiny$\leftrightarrow$}}{V}(\textbf{r}_2,\textbf{r}^\prime;\omega) + \cdots
\end{multline}
Establishing the above most general structure of the scattering transition matrix, we can discuss the conditions for reciprocity. As before, reciprocity of a scattering process from a plane wave of wavevector $\textbf{k}$ and polarization $\sigma$ to a wavevector $\textbf{k}^\prime$ and polarization $\sigma^\prime$ requires
\be
\langle \textbf{k},\sigma |\hat{T}(\omega) | \textbf{k}^\prime,\sigma^\prime \rangle = \langle -\textbf{k}^\prime,\sigma^\prime  |\hat{T}(\omega) |- \textbf{k},\sigma \rangle
\label{reciprocity}
\ee
Note that the scattering theory we have been using, including the calculation of transition amplitudes, is all done in the ``far field" for which standard scattering theory is developed and $\omega = ck = ck^\prime$ is applied. Recently generalization to near field have been demonstrated for some situations \cite{Carminati1998}.

\section{Kerr effect from a clean, single-band chiral superconductor}

Using the example of a $p_x \pm ip_y $ superconductor, it is easy to see that while indeed an asymmetry in the optical conductivity may exist, the resulting Kerr effect of a single band, pure system (that is, translational symmetry is respected) is identically zero \cite{Roy2008,Lutchyn2008}. To show this property, we follow a simple approach also mentioned by Lutchyn {\it et al.} \cite{Lutchyn2008}. For simplicity, let us consider a $p_x \pm ip_y $ superconductor. If each Cooper pair carries angular momentum $\hbar$ in the $\hat{z}$-direction, the average angular momentum carried by the condensate is $\langle L_z\rangle =\hbar N/2$, where $N$ is the electron density. The respective orbital magnetization is then
\be
\textbf{M}  = \frac{-e}{2mc}\langle L_z\rangle = \frac{-eN\hbar}{4mc}\hat{z}
\ee
where $m$ is the electron mass and  $\rho$ is the electron charge density.  The anomalous current associated with this magnetization is then \cite{Mermin1980,Balatskii1985}
\be
\textbf{j}^{a}=c \nabla \times\textbf{M} = \frac{\hbar e}{4m}\hat{z}\times \nabla N= \frac{\hbar e}{4m}\frac{\partial N}{\partial \mu} \hat{z}\times \nabla \mu
\ee
where $\mu$ is the chemical potential. This expression can also be written in the form
\be
\textbf{j}^{a}= \frac{e^2}{h}\left( 2\pi \frac{\hbar^2}{2m}\frac{\partial N}{\partial \mu} \right) \hat{z}\times \nabla \left(\frac{\mu}{2e}\right)\equiv \sigma_{xy} \hat{z}\times \nabla \left(\frac{\mu}{2e}\right)
\ee
where, following ref- \cite{Mineev2007}, we identified the coefficient with the off diagonal conductivity $\sigma_{xy}$.  It is easy to see now that in general
\be
 \nabla \mu= \nabla \left( 2eV + \hbar\frac{\partial \varphi}{\partial t}\right )
 \ee
where $\nabla V = -\textbf{E}$ is the actual electric field, and 
\be
\varphi = \phi-\frac{2e}{\hbar c} \int\textbf{A} \cdot d {\bf \ell}
\ee
is the total phase change along the gradient.

The final result is therefore
\be
\textbf{j}^{a}= \sigma_{xy} \hat{z}\times \nabla \left[ -\textbf{E}  +\frac{\partial}{\partial t} \left(\frac{\hbar}{2e}\nabla \phi -\frac{1}{c}\textbf{A} \right) \right] \equiv 0
\label{anomalous}
\ee
where the zero final value originates from the London equation for the supercurrent $\textbf{j}_s$
\be
(2m)\frac{\partial \textbf{j}_s}{\partial t}=(2e)\rho_s\textbf{E}
\ee
applied to the supercurrent  $\textbf{j}_s =(e\rho_s/2m)[\nabla \phi -(2e)/c \textbf{A}]$ (where  $\rho_s$ is the superfluid density.) Since the anomalous current vanishes, the Kerr effect is identically zero for this case of clean, single-band superconductors.


\section*{Acknowledgements}
These notes are the result of numerous discussions with students and colleagues on the possible detection of Kerr effect in a variety of materials. I am particularly grateful to discussions with Alexander Fried, as well as Weejee Cho, Steven Kivelson, Sri Raghu, Elizabeth Schemm, Chandra Varma, Victor Yakovenko, and Jing Xia. This work was supported by the Department of Energy Grant  DE-AC02-76SF00515.\\

\section*{References}

\end{document}